\documentclass[12pt]{iopart}

\usepackage{graphicx}
\usepackage{dcolumn}
\usepackage{bm}

\usepackage{color} 

\begin{document}

\title{Chirality-Controlled Enantiopure Crystal Growth of a Transition Metal Monosilicide by a Floating Zone Method}

\author{Yusuke Kousaka$^1$, Satoshi Iwasaki$^2$, Taisei Sayo$^1$, Hiroshi Tanida$^3$, Takeshi Matsumura$^4$, Shingo Araki$^5$, Jun Akimitsu$^2$, Yoshihiko Togawa$^1$}
\address{$^1$ Department of Physics and Electronics, Osaka Prefecture University, Sakai, Osaka 599-8531, Japan}
\address{$^2$ Research Institute for Interdisciplinary Science, Okayama University, Okayama, Okayama 700-8530, Japan}
\address{$^3$ Liberal Arts and Sciences, Toyama Prefectural University, Imizu, Toyama 939-0398, Japan}
\address{$^4$ Graduate School of Advanced Sciences of Matter, Hiroshima University, Higashihiroshima, Hiroshima 739-8530, Japan}
\address{$^5$ Department of Physics, Faculty of Science, Okayama University, Okayama, Okayama 700-8530, Japan}

\ead{koyu@pe.osakafu-u.ac.jp}

\vspace{10pt}

\begin{abstract}
We performed a crystal growth to obtain chirality-controlled enantiopure crystals using a laser-diode-heated floating zone (LDFZ) method with a composition-gradient feed rod. 
It has been argued that the crystal handedness of $T$Si ($T$ : transition metal) is fixed depending on $T$ in the case of the ones grown by the conventional methods.
We found that right-handed single crystals of CoSi and MnSi were grown from
the composition gradient feed rods that consist of FeSi--CoSi and FeSi--MnSi, respectively.
The obtained CoSi and MnSi crystals inherit the chirality from the seed part of FeSi, which grows in a right-handed structure, and thus have the chirality opposite to that for the crystals in the literature.
The LDFZ method with the feed rods with various combinations of $T$Si compounds enables a flexible control of the chirality of $T$Si and will be useful for clarifying the interplay between the crystalline chirality and chirality-induced physical responses.
\end{abstract}

%
%
%
%
%

\section{Introduction}

Chirality, which represents handedness of the objects, is one of the most fundamental characteristics in nature.
In chiral helimagnetism, it is important to understand the interplay between crystallographic and magnetic chiralities \cite{Kishine2015,Togawa2016}
because properties of a helical magnetic structure strongly depend on the chiral skeleton structure that could allow an antisymmetric Dzyaloshinskii-Moriya (DM) interaction \cite{Dzyaloshinskii1958, Moriya1960}.

In inorganic chiral compounds, it is still a challenge to control the crystallographic chirality of materials.
It is difficult to prevent inorganic materials from forming racemic-twinned crystalline grains of left- and right-handedness during a crystal growth.
One possible method of controlling the crystallographic chirality is an application of spontaneous crystallization process in a stirring solution, which enables a growth of crystals of single handedness \cite{Kondepudi1990}.
For example, a chiral magnetic compound CsCuCl$_{3}$, which is soluble in water and forms racemic-twinned crystals in conventional crystallization methods \cite{Ohsumi2013, Kousaka2009}, grows to be an enantiopure crystal of millimeter in size in the stirring solution \cite{Kousaka2014}. Such an enantiopure crystal can be utilized as a seed crystal to obtain a centimeter-sized crystal of the selected handedness \cite{Kousaka2017, Nakamura2020}.
Although the stirring method is applicable only to soluble materials, this study enlightens the importance of a usage of enantiopure seed crystals in the synthesis of mono-chiral crystals.

Intermetallic compounds $T$Si ($T:$ transition metal) with a B20 type chiral crystal structure are one of the representative chiral magnetic compounds investigated in the community because of the formation of Skyrmion lattice \cite{Kadowaki1982,Bogdanov1989,Muhlbauer2009}.
It is known that, while polycrystalline $T$Si samples consist of racemic-twinned crystalline grains \cite{TSi1}, single crystals of $T$Si can be obtained with the single handedness  \cite{TSi2,Ishida1985,TSi3,TSi4}.
The crystallographic chirality depends on the elements of $T$ site.
The right-handed crystals are stabilized in $T =$ Fe, while the left-handed ones are grown in $T =$ Mn and Co.
In the case of Fe$_{1-x}$Co$_{x}$Si, the crystal chirality remains right-handed up to $x = 0.15$, where Fe is rich. However, it flips to left-handed when $x$ is larger than 0.20 \cite{TSi3}.
Although it is unknown why single crystals of $T$Si stabilize with only a particular handedness, it would be interesting to obtain $T$Si crystals with the chirality opposite to that of the crystals grown by the conventional methods such as a Czochralski (Cz) process.

In this paper, we report the chirality control of $T$Si single crystals by using a laser-diode-heated floating zone (LDFZ) method together with a feed rod of $T$Si with a composition gradient of $T$ along the growth direction. Indeed, we grew $T$Si single crystals stabilized in a crystal structure with the handedness opposite to that of the ones grown by the conventional method.
Such chirality-controlled crystals can be utilized as a seed crystal for the successive crystal growth. 
Experiments with the enantiopure crystal with the fixed handedness will contribute to revealing the interplay between crystallographic and magnetic chiralities.

\section{Crystal structure and its chirality}

\begin{figure}[tb]
\begin{center}
\includegraphics[width=10cm]{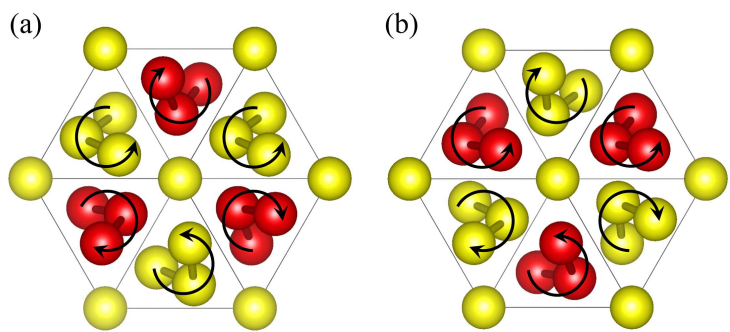}
\end{center}
\caption{Crystal structures of (a) right-handed and (b) left-handed $T$Si ($T:$ transition metal), viewed along the $<111>$. Red and yellow balls represent $T$ and Si atoms, respectively.
The black arrows are an eye guide to recognize the sense of screw alignments of the $T$ and Si atoms located at Wyckoff $4a$ positions.
}
\label{f-TSi-str}
\end{figure}

The crystal structures of $T$Si belong to the space group $P2_{1}3$, which allows the right- and left-handed crystal structures.
When the structures are viewed along the $<$111$>$, $T$ and Si chains are twisted right- or left-handed, as shown in Fig.~\ref{f-TSi-str}.
Right-handed screws of Si atoms and left-handed ones of $T$ atoms are found in the right-handed crystal structures and vice versa in the left-handed crystal.
Here, the handedness of the crystal structures follows the definition given in the literature \cite{TSi2}, where the handedness of the crystal is determined by the Si helices.

$T$ and Si atoms occupy Wyckoff $4a$ positions, as expressed by $(u,u,u)$ and its equivalent positions in $P2_{1}3$.
The handedness of the helices depends on whether a value of $u$ is smaller or larger than 0.5.
An inversion operation switches the chirality of the crystal, giving the equivalent atomic position of $(1-u,1-u,1-u)$ for $4a$.
Absolute structure analysis using X-ray shows that the values of $u$ take about $0.14 \textendash 0.16$ for Si and $0.84 \textendash 0.86$ for $T$ in the right-handed crystals, while about $0.84 \textendash 0.86$ for Si and $0.14 \textendash 0.16$ for $T$ in the left-handed crystal \cite{TSi3,Ishikawa1976,Ishikawa1977}.
These values indeed satisfy the relationship of $u$ with respect to the handedness.

\section{Experimental methods}

To prepare the feed and seed rods used for the LDFZ method, the polycrystalline samples of $T$Si were synthesized by an arc melting method.
A stoichiometric mixture of $T$ and Si in the molar ratio of 1 : 1 was melted in an Ar atmosphere.
The obtained ingots were pulverized to powders and then the rod-shaped polycrystals were prepared by sintering the powders.
Single crystals of $T$Si were obtained by the LDFZ furnace (equipped with five 200 W lasers, Crystal Systems Corporation).
The crystal growth was performed under an Ar atmosphere, and the Ar flowing rate was about 3 $l$$/$min.
To grow single crystals, traveling speeds of the feed and seed rods were set to be the same at 5--10 mm$/$h.
To obtain a stable molten zone, the feed and seed rods were rotated in the opposite directions at 4 and 5 rpm, respectively.

The floating zone (FZ) method is known as a technique to obtain single crystals of centimeters in length and several millimeters in diameter.
The feed and seed rods for the FZ crystal growth generally consist of the same compound.
However, in some compounds, it is difficult to grow single crystals due to technical reasons such as
a very slow speed of the crystal growth and instability of the molten zone.
These problems are sometimes solved by using a composition-gradient feed rod.

In the case of superconducting copper oxide compounds La$_{2-x}$Sr$_{x}$CuO$_{4}$ (La214) \cite{Ikeuchi2003},
a successive growth of single crystals with different concentrations was performed by using the seed rod of single crystal made at a particular value of the Sr concentration
and the feed rod, which had a step-like Sr-concentration gradient along the rod. Eventually, the obtained single crystal had a gradient of Sr-concentration along the grown crystal.
In the case of Bi$_{2+x}$Sr$_{2-x}$CuO$_{6-\delta}$ \cite{Luo2007, Enoki2010, Enoki2013},
it is rather hard to be crystallized when the Sr amount is large. To obtain single crystals of Sr-rich concentration, the feed rod of gradient $x$ concentration was prepared.
More precisely, the seed rod with Bi-rich composition was used for stabilizing the crystal growth at the initial stage. After the connection of the feed rod to the seed rod and following crystal growth, the single crystal was obtained with a gradual change of the composition from Bi-rich to Sr-rich phases since the crystal growth even with Sr-rich composition was stabilized.
Consequently, the seed rod kept the mono-domain and the same crystallographic orientation along the rod. This method enables the growth of single crystals of Bi$_{2+x}$Sr$_{2-x}$CuO$_{6-\delta}$ with various $x$ values \cite{Enoki2010}.

These studies indicate that the crystal growth, initiated with the feasible composition and followed by the growth with the composition of unstable regime, would be useful for stabilizing the crystal growth.
However, these compounds form centrosymmetric crystal structures. It is still unknown whether the crystal growth with composition-gradient feed rods inherits chiral crystal structures.
In this study, we have tried to use the crystal growth method with composition-gradient feed rods in synthesizing chiral magnetic crystals with the controlled chirality.

The crystallographic chirality of $T$Si was evaluated by X-ray absolute structure analysis using anomalous scattering.
In a non-centrosymmetric system including a chiral system,
the intensities of a pair of reflections at $(h, k, l)$ and $(\bar{h}, \bar{k}, \bar{l})$, termed Bijvoet pairs \cite{Bijvoet1954}, are not equivalent.
One can obtain the Flack parameter $x$~\cite{Flack1983, Berardinelli1985, Flack1999, Parson2013}
as a fitting parameter of the structure analysis, which is expressed as $I_{obs}(h,k,l)=(1-x)I_{cal}(h, k, l)+xI_{cal}(\bar{h}, \bar{k}, \bar{l})$. Here, $I_{obs}$ and $I_{cal}$ are the observed and calculated intensities, respectively.
The $I_{cal}(h,k,l)$ and $I_{cal}(\bar{h}, \bar{k}, \bar{l})$ are calculated from the assumed absolute structure in the analysis and the inverted one, respectively.
Therefore, if the refined Flack parameter $x$ is 1, the assumed chiral crystal structure has to be inverted into that with the opposite handedness.
If the refined Flack parameter $x$ is 0, the assumed chiral crystal structure is correct.

To evaluate the crystallographic chirality of single crystals of $T$Si obtained by the FZ method, we cut some small portions, and examined them by means of X-ray oscillation photograph.
The data were collected using Rigaku R-AXIS RAPID and Bruker D8-QUEST with Mo $K\alpha$ radiation at room temperature.
With applying an empirical absorption correction to the observed intensity, the absolute structures were solved by the direct method and refined using the SHELXL software package \cite{Shelx}.
The chiral crystal structure was determined by the Flack parameter $x$, which was obtained as one of the refined parameters in the analysis.
In the case of the right-handed structure, the Flack parameter $x$ shows 0 when the values of $u$ for Wyckoff $4a$ positions are about $0.14 \textendash 0.16$ for Si and $0.84 \textendash 0.86$ for $T$.
In the case of the left-handed structure, the Flack parameter $x$ shows 0 when the values of $u$ are about $0.84 \textendash 0.86$ for Si and $0.14 \textendash 0.16$ for $T$.

\section{Experimental results and discussion}

First, the crystal syntheses were performed with the seed and feed rods consisting of the single compound
of B20 materials. The single crystals of FeSi, CoSi, and MnSi were successfully obtained by the LDFZ method.
The results of the absolute structure analysis of these crystals are summarized in Table \ref{tbl-FeCoMnSi-conventional}.
The lattice constant $a$ of FeSi is larger or smaller than that of CoSi and MnSi below 2 percent, respectively.
It is clear that the lattice constant $a$ reduces with an increase of the atomic number of $T$.
Namely, the value of $a$ is proportional to the atomic radius of $T$ and can be a criterion on determining $T$ atoms contained in the synthesized crystals.
The handedness of the crystals was right-handed for FeSi and left-handed for CoSi and MnSi. These results are consistent with those reported in the literature \cite{TSi2, TSi3, TSi4}.

\begin{table}[htbp]
      \caption{Crystal structure analysis of $T$Si, obtained by the LDFZ method using the feed rod with the single composition of $T$Si.
                    Note that $T$ and Si atoms are located at Wyckoff sites $4a$ $(u_{T},u_{T},u_{T})$ and $(u_{\rm{Si}},u_{\rm{Si}},u_{\rm{Si}})$, respectively.
                   }
      \label{tbl-FeCoMnSi-conventional}

  \begin{center}
  
      \begin{tabular}{rrrr}
      \hline
      \multicolumn{1}{c}{} & \multicolumn{1}{c}{FeSi} & \multicolumn{1}{c}{CoSi} & \multicolumn{1}{c}{MnSi}\\
      \hline
      \multicolumn{1}{c}{Space group} & \multicolumn{3}{c}{$P2_{1}3$}\\
      \multicolumn{1}{c}{$a$ [\AA]} & \multicolumn{1}{c}{$4.4865(5)$} & \multicolumn{1}{c}{4.4450(6)} & \multicolumn{1}{c}{4.575(5)}\\
      \multicolumn{1}{c}{Fomula units} & \multicolumn{1}{c}{4} & \multicolumn{1}{c}{4} & \multicolumn{1}{c}{4}\\
      \multicolumn{4}{c}{}\\
      \multicolumn{1}{c}{Atomic positions} & \multicolumn{3}{c}{}\\
      \multicolumn{1}{c}{$u_{T}$} & \multicolumn{1}{c}{$0.8635(1)$} & \multicolumn{1}{c}{$0.1434(1)$} & \multicolumn{1}{c}{$0.1367(1)$}\\
      \multicolumn{1}{c}{$u_{\rm{Si}}$} & \multicolumn{1}{c}{$0.1574(2)$} & \multicolumn{1}{c}{$0.8436(3)$} & \multicolumn{1}{c}{$0.8454(2)$}\\
      \multicolumn{4}{c}{}\\ 
      \multicolumn{1}{c}{$R1$} & \multicolumn{1}{c}{$0.0127$} & \multicolumn{1}{c}{0.0247} & \multicolumn{1}{c}{0.0143}\\
      \multicolumn{1}{c}{$wR2$} & \multicolumn{1}{c}{$0.0308$} & \multicolumn{1}{c}{0.0651} & \multicolumn{1}{c}{0.0456}\\
      \multicolumn{1}{c}{Flack parameter $x$} & \multicolumn{1}{c}{$0.011(25)$} & \multicolumn{1}{c}{$-0.012(66)$} & \multicolumn{1}{c}{$0.043(36)$}\\
      \multicolumn{1}{c}{Absolute structure} & \multicolumn{1}{c}{Right-handed} & \multicolumn{1}{c}{Left-handed} & \multicolumn{1}{c}{Left-handed}\\
      \hline
      \end{tabular}

  \end {center}
\end{table}

\begin{figure*}[tb]
\begin{center}
\includegraphics[width=15.0cm]{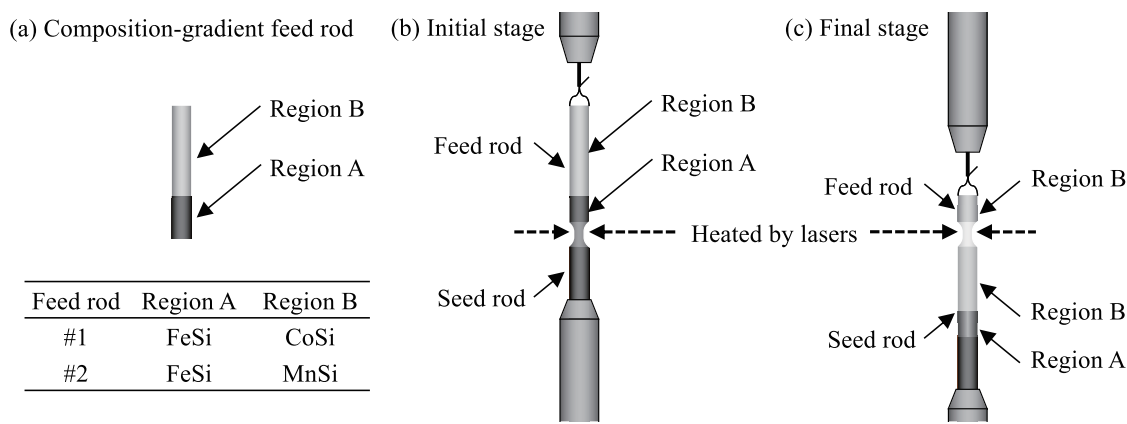}
\end{center}
\caption{
Schematic drawings of a crystal growth using laser-diode-heated floating zone (LDFZ) method with a composition-gradient feed rod.
(a) The composition-gradient feed rod consists two regions with different compounds.
The table in (a) indicates a combination of the $T$Si compounds used in the present study.
(b) and (c) The LDFZ procedure.
The crystal growth starts with the region A and ends with the region B preferring the chirality opposite to that of FeSi.
The combination of B20 materials, as shown in the table in (a), enables the control of chirality of the grown crystals.
}
\label{f-FZ}
\end{figure*}

To obtain the single crystals of $T$Si with the controlled chirality, we performed the LDFZ method with composition-gradient feed rods. In the present study, the composition-gradient feed rods consist of two phases of $T$Si, as shown in Fig.~\ref{f-FZ}(a).
Note that a drastic change of the melting temperature occurs at the boundary between the regions A and B, which requires a precise control of the laser power to keep the molten zone stabilized by the well-focused laser.
This is one of the advantages of the crystal growth using the LDFZ method.
If the crystallographic information of the region A succeeds to that of the region B during the growth,
the crystal at the region B can present right-handed chirality, which is hardly obtained by the conventional LDFZ method.

The first experiment was performed with the composition-gradient feed rod \#1 shown in the table in Fig.~\ref{f-FZ}(a), consisting of FeSi and CoSi in the regions A and B, respectively.
We aimed to obtain a CoSi single crystal of right-handed, which is opposite to the handedness of the CoSi shown in Table~\ref{tbl-FeCoMnSi-conventional}.
In this case, the feed rod consists of FeSi and CoSi. The growth starts with FeSi in the region A, as shown in Fig.~\ref{f-FZ}(b).
The seed rod is initially crystallized into the single domain of FeSi. Then, the phase of the feed rod changes into CoSi at the beginning of the region B.
In the successive crystal growth, as shown in Fig.~\ref{f-FZ}(c), CoSi is crystallized on the seed crystal of FeSi.

\begin{figure}[tb]
\begin{center}
\includegraphics[width=10cm]{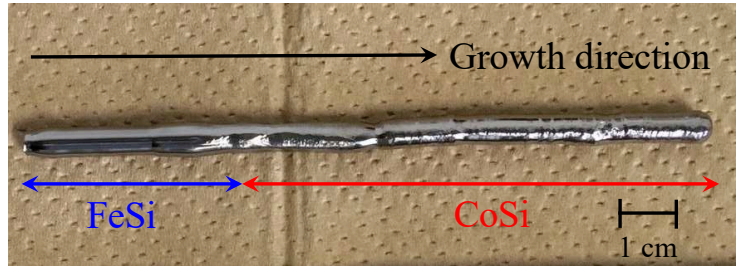}
\end{center}
\caption{An optical photograph of the single crystal of FeSi and CoSi, obtained by the LDFZ method with a composition-gradient feed rod.}
\label{f-TSi-photo}
\end{figure}

Figure~\ref{f-TSi-photo} shows an optical photo of the obtained single crystal of FeSi and CoSi.
Note that the crystalline quality of the single crystal was evaluated by Laue X-ray diffraction with an imaging plate.
Both FeSi and CoSi regions form a single crystalline domain and have the same crystallographic orientation.
Therefore, the crystallographic orientation of FeSi succeeds to that of CoSi.
As shown in the following, it is found that the crystallographic chirality of the CoSi region is also the same as those of the seed rod of FeSi, obtained at the initial stage of the growth.

To evaluate the crystal chirality of the single crystal grown by the composition-gradient feed rod consisting of FeSi and CoSi, some small portions were cut at both the regions of FeSi and CoSi.
The crystal chirality of these portions was determined by the deducted Flack parameter $x$ in the absolute structure analysis using X-ray oscillation photograph.
The results are summarized in Table~\ref{tbl-CoMnSi-gradient}.
Note that the refined Flack parameters of all the portions of $T$Si are the same within three-sigma error bar.
The Flack parameter $x$ of CoSi is almost zero with the parameters of the atomic position $u_{T}$ and $u_{\rm{Si}}$ being approximately 0.86 and 0.16, respectively.
As discussed in the literature \cite{TSi2}, the combination of the parameters indicates the right-handed crystal structure.
Importantly, the lattice constant in the CoSi region reproduces the value obtained by the conventional FZ method, shown in Table~\ref{tbl-FeCoMnSi-conventional}, while the crystal chirality of CoSi turns to be the same as that of FeSi.
Therefore, it is concluded that the obtained CoSi crystal exhibits the same chirality as that of the seed FeSi crystal.
Namely, the single crystal of right-handed CoSi, in which the handedness is opposite to that of the crystals by the conventional methods, was successfully obtained.
The first example mentioned above indicates that the LDFZ crystal growth using the composition-gradient feed rod enables us to obtain single crystals with a different combination of B20 compounds and control the chirality of the crystals.

The second example is the single crystal growth of MnSi with the composition-gradient feed rod \#2 shown in the table in Fig.~\ref{f-FZ}(a).
Similar results were observed for the case of the feed rod with FeSi and MnSi. As indicated in Table~\ref{tbl-CoMnSi-gradient}, the absolute structure analysis shows that the right-handed crystal structure was obtained in the MnSi region.
While the lattice constant $a$ of MnSi and CoSi is different from that of FeSi below 2 percent, the single crystals of CoSi and MnSi inherit the crystallographic chirality from those of FeSi.
The LDFZ method with the composition-gradient feed rods can obtain chiral single crystals of the handedness different from that of the crystals by the conventional methods.

\begin{table}[htbp]
      \caption{Crystal structure analysis of $T$Si, obtained by the LDFZ method with the composition-gradient feed rod.
                    It consists of FeSi at the region A and $T$Si at the region B as shown in Fig.~\ref{f-FZ}(a).}
      \label{tbl-CoMnSi-gradient}

  \begin{center}

      \begin{tabular}{rrr}
      \hline
      \multicolumn{1}{c}{} & \multicolumn{1}{c}{CoSi} & \multicolumn{1}{c}{MnSi}\\
      \hline
      \multicolumn{1}{c}{Space group} & \multicolumn{2}{c}{$P2_{1}3$}\\
      \multicolumn{1}{c}{$a$ [\AA]} & \multicolumn{1}{c}{$4.4457(5)$} & \multicolumn{1}{c}{$4.5667(6)$}\\
      \multicolumn{1}{c}{Fomula units} & \multicolumn{1}{c}{4} & \multicolumn{1}{c}{4}\\
      \multicolumn{3}{c}{}\\
      \multicolumn{1}{c}{Atomic positions} & \multicolumn{2}{c}{}\\
      \multicolumn{1}{c}{$u_{T}$} & \multicolumn{1}{c}{$0.8565(1)$} & \multicolumn{1}{c}{$0.8631(1)$}\\
      \multicolumn{1}{c}{$u_{\rm{Si}}$} & \multicolumn{1}{c}{$0.1563(2)$} & \multicolumn{1}{c}{$0.1544(1)$}\\
      \multicolumn{3}{c}{}\\ 
      \multicolumn{1}{c}{$R1$} & \multicolumn{1}{c}{$0.0145$} & \multicolumn{1}{c}{0.0072}\\
      \multicolumn{1}{c}{$wR2$} & \multicolumn{1}{c}{$0.0305$} & \multicolumn{1}{c}{0.0178}\\
      \multicolumn{1}{c}{Flack parameter $x$} & \multicolumn{1}{c}{$-0.044(25)$} & \multicolumn{1}{c}{$-0.008(15)$}\\
      \multicolumn{1}{c}{Absolute structure} & \multicolumn{1}{c}{Right-handed} & \multicolumn{1}{c}{Right-handed}\\
      \hline
      \end{tabular}

  \end {center}
\end{table}

\section{Conclusion}

In summary, we successfully grew enantiopure single crystals of FeSi, CoSi, and MnSi using the LDFZ method.
Monosilicide compounds $T$Si are known as an example to form only left- or right-handed crystal domain, and the chirality depends on the elements of $T$ site \cite{TSi2,Ishida1985,TSi3,TSi4}. Indeed, we confirmed that in the case of the conventional LDFZ growth using the seed and feed rods consisting of the same compounds, the handedness of the obtained crystals was consistent with the reported ones. However, when using the composition-gradient feed rods, the handedness of the obtained crystals was opposite to the reported one, which is hardly synthesized by the conventional crystal growth methods.
These features indicate that the obtained crystals inherit the chirality from the seed crystals.

In inorganic chiral compounds, enantiopure crystal growth has been mainly performed in a spontaneous crystallization from a stirring water solution. However, such a technique is not applicable in most of inorganic compounds because they are not soluble in water.
On the other hand, the present LDFZ technique holds a molten zone at high temperature just above the melting temperature of the compound.
It can crystallize the domain with a desired handedness by inheriting the crystallographic orientation and chirality from the seed crystal. In this sense, this technique may be applicable in many kinds of inorganic chiral compounds.

It is still unclear why the $T$Si compounds spontaneously form the crystalline structure with the single handedness.
Nevertheless, the present study revealed the importance of a usage of the seed crystal with the controlled chirality. Similar technique has been already demonstrated in the Cz process with an enantiopure seed crystal using a tri-arc furnace \cite{TSi5}.
However, the Cz method sometimes fails to succeed chirality from the seed crystal because of growth conditions used in the tri-arc furnace.
The growth temperature of the tri-arc furnace is inevitably much higher than the melting point of the compounds.
Under such harsh conditions, the chirality of the seed crystal may be destroyed when the mono-chiral seed crystal is put into the liquid melt in the initial process of crystal growth.

On the other hand, in the LDFZ method, as explained above, the precise temperature control is feasible in order to keep the molten zone stabilized by the well-focused laser. This is likely to work well in inheriting the crystallographic information of the seed crystal into the connection region. Furthermore, the sequent crystal growth using the composition-gradient feed rod keeps the crystallographic information between the regions A and B consisting of different compounds.
Therefore, the advantages of the LDFZ method can be found in flexible choice of material combination and chirality as well as the size of the grown crystals.
The chirality-controlled enatiopure crystals obtained by the present LDFZ method will be useful for clarifying the interplay between the crystalline chirality and chirality-induced physical responses such as magnetochiral effect \cite{Yokouchi2017,Nakagawa2017,Aoki2019} and chirality-induced spin-selectivity (CISS) effect  recently found in inorganic chiral crystals \cite{Inui2020,Shiota2021,Shishido2021}.

\ack

We express our sincere appreciation to M. Enoki for fruitful discussion.
This work was supported by JSPS KAKENHI Grant Numbers 15H05885, 17H06137, 19KK0070 and 20H02642.

\end{document}